\pdfoutput=1
\documentclass[12pt,letterpaper,floatfix]{article}
\usepackage{amsmath}
\usepackage{amssymb}
\usepackage{epsfig}
\usepackage{euscript}
\usepackage{setspace}
\def\mathswitchr#1{\relax\ifmmode{\mathrm{#1}}\else$\mathrm{#1}$\fi}

\def\rQCED{{\rm QCED}}

\newcommand {\pslash}{\hbox{$\not\hbox{\kern-2.3pt $p$}$}}

\newcommand{\FYFS}{F_{\mathrm{YFS}}}
\usepackage{cite}

\usepackage{epic}
\def\alf1{ {\alpha\over\pi} }
\def\rQCED{{\rm QCED}}
\begin{document}
%\input{feynman} 
%=======================================================================
%\begin{titlepage}
%\begin{flushright}
%{\bf CERN-PH-TH/2011-077}\\
%{\bf BU-HEPP-11-04}\\
%{\bf Dec., 2011}\\
%\end{flushright}
%\vspace{0.05cm}
\begin{titlepage}
\begin{center}
{\bf \large New Approach to Hard Corrections in Precision QCD for LHC and FCC Physics}\\
%\end{center}
%\ShortTitle{Exact Amplitude-Based Resummation in Quantum Field Theory: Recent Results}
\vspace{2mm}
%\begin{center}
%%  {\bf   S. Jadach$^{a,b}$ and B.F.L. Ward$^{c,d}$}
%\author{B.F.L. Ward\\%
%    \thanks{Work supported in part by D.o.E. grant DE-FG02-09ER41600.}\\
%      Baylor University\\
%        E-mail:\email{bfl\_ward@baylor.edu}}
%    S.K. Majhi
% \footnote{Work supported by 
%grant Pool No. 8545-A, CSIR, IN.}\\
%      Indian Association for the Cultivation of Science, Kolkata, India\\
%        E-mail: tpskm@iacs.res.in\\
%A. Mukhopadhyay\\%
%    \thanks{Work supported in part by D.o.E. grant DE-FG02-09ER41600.}\\
%      Baylor University, Waco, TX, USA\\
%        E-mail: aditi\_mukhopadhyay@baylor.edu\\
B.F.L. Ward\\%
%    \thanks{Work supported in part by D.o.E. grant DE-FG02-09ER41600.}\\
      Baylor University, Waco, TX, USA\\
        E-mail: bfl\_ward@baylor.edu\\
%S.A. Yost
% \footnote{Work supported in part by U.S.
%D.o.E. grant DE-FG02-10ER41694 and grants from The Citadel Foundation.}\\
%      The Citadel, Charleston, SC, USA\\
%        E-mail: scott.yost@citadel.edu\\
\end{center}
\vspace{2mm}
\centerline{\bf Abstract}
We present a new approach to the realization of hard fixed-order corrections in predictions for the processes probed in high energy colliding hadron beam devices, with some emphasis on the LHC and the future FCC devices. We show that the usual unphysical divergence of such corrections as one approaches the soft limit is removed in our approach, so that
we would render the standard results to be closer to the observed exclusive distributions. We use the single $Z/\gamma*$ production and decay to lepton pairs as our prototypical example, but we stress that the approach has general applicability. In this way, we open another part of the way to rigorous baselines for the determination of the theoretical precision tags for LHC physics, with an obvious generalization to the future FCC as well. \\
\vspace{1cm}
\begin{center}
 BU-HEPP-14-03, June, 2014\\
\end{center}

%          BU-HEPP-11-04,\\ 
%             Dec., 2011 }
%\end{center}
\end{titlepage}
%\begin{document}
%\\
%\vskip 3mm
%\centerline{Invited talk presented by B.F.L. Ward at RADCOR 2011, Chennai, India}
%\vskip 16mm
% 
%\vspace{10mm}
%\renewcommand{\baselinestretch}{0.1}
%\footnoterule
%\noindent
%{\footnotesize
%\begin{itemize}
%\item[${\dagger}$]
%Work partly supported by US DOE grant DE-FG02-09ER41600. 
% the Polish Government
%grants KBN 2P30225206 and 2P03B17210, the Maria Sk\l{}odowska-Curie
%Joint Fund II PAA/DOE-97-316, and
%by NATO Grant PST.CLG.980342.
%, and by
%Polish Government grant 5P03B09320.
%\end{itemize}
%}
%\vspace{0.5cm}
%\begin{flushleft}
%{\bf UTHEP-00-0101}\\
%{\bf Jan, 2000}\\
%\end{flushleft}

%\end{titlepage}
 
%\baselineskip=11pt 
%=======================================================================
\def\Kmax{K_{\rm max}}\def\ieps{{i\epsilon}}\def\rQCD{{\rm QCD}}
%\renewcommand{\theequation}{\arabic{equation}}
%\font\fortssbx=cmssbx10 scaled \magstep2
%\renewcommand\thepage{}
%\vfill\eject
%\parskip.1truein\parindent=20pt\pagenumbering{arabic}\par

%\section{\bf Introduction}\par

\begin{figure}[h]
\begin{center}
%x\epsfig{file=pent-1.eps,width=140mm}
\setlength{\unitlength}{0.1mm}
\begin{picture}(1600, 930)
\put( 810, 920){\makebox(0,0)[cb]{\bf\Large $Z/\gamma^*$ transverse momentum $\left(d\sigma/d\phi^*_\eta(\ell\ell)\right)$} }
\put(  40, -20){\makebox(0,0)[lb]{\includegraphics[width=150mm]{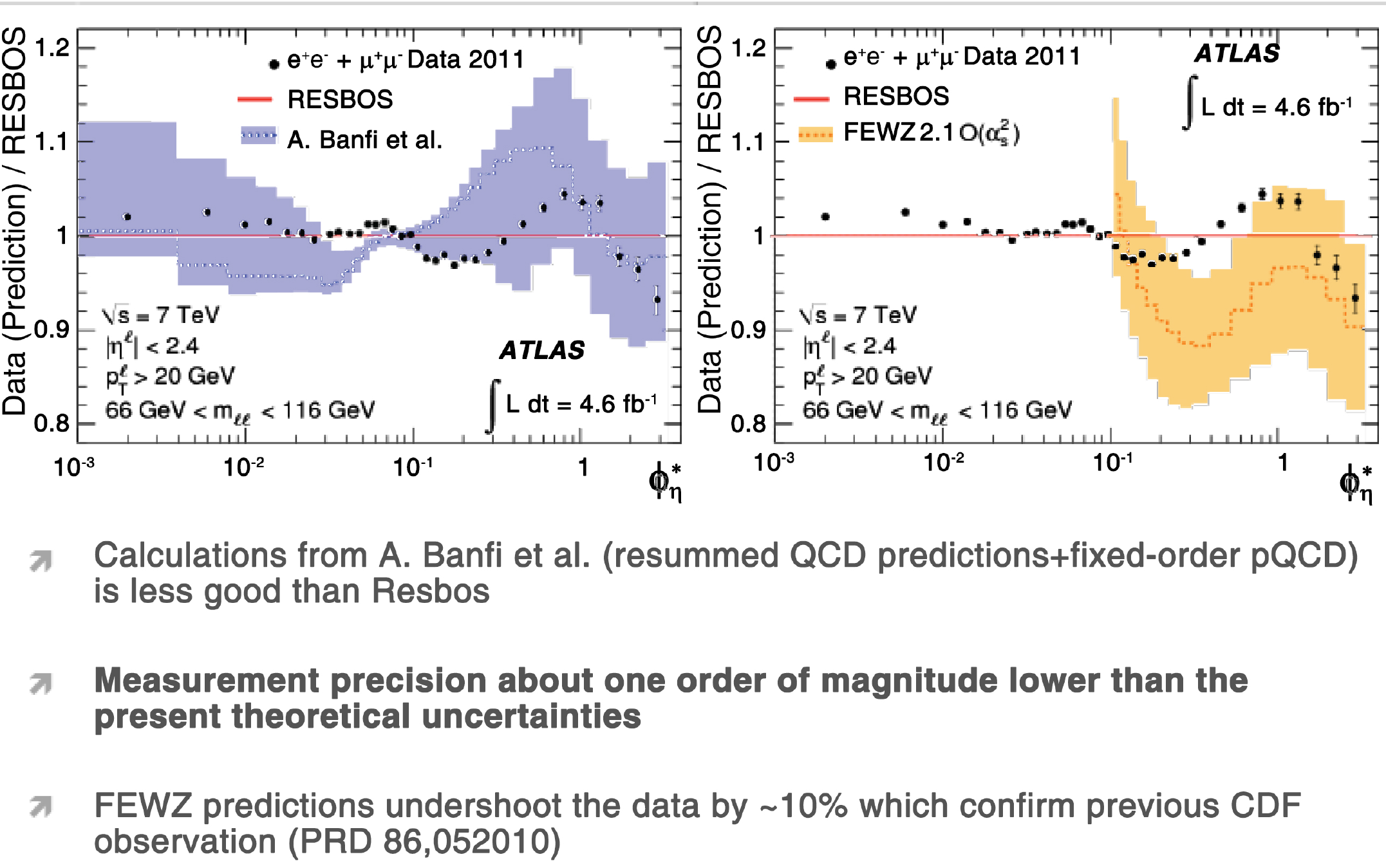}}}
\end{picture}
\end{center}
\caption{\baselineskip=11pt Comparisons of some theoretical predictions with the ATLAS  $Z/\gamma^*$ $\phi^*_\eta$ spectrum in single $Z/\gamma^*$ production with decay to lepton pairs as given in Ref.~\cite{atlas-hassani}. Here Banfi {\em et al.} refers also to a resummed calculation of the ``CSS'' type~\cite{css}, so that it has the same physical precision limitations as RESBOS~\cite{resbos1,resbos2,resbos3} as discussed in Ref.~\cite{1305-0023} -- see the second reference in Refs.~\cite{phietastr}. }
\label{atlas-hass-1}
\end{figure}

\begin{figure}[h]
\begin{center}
%x\epsfig{file=pent-1.eps,width=140mm}
\includegraphics[width=100mm]{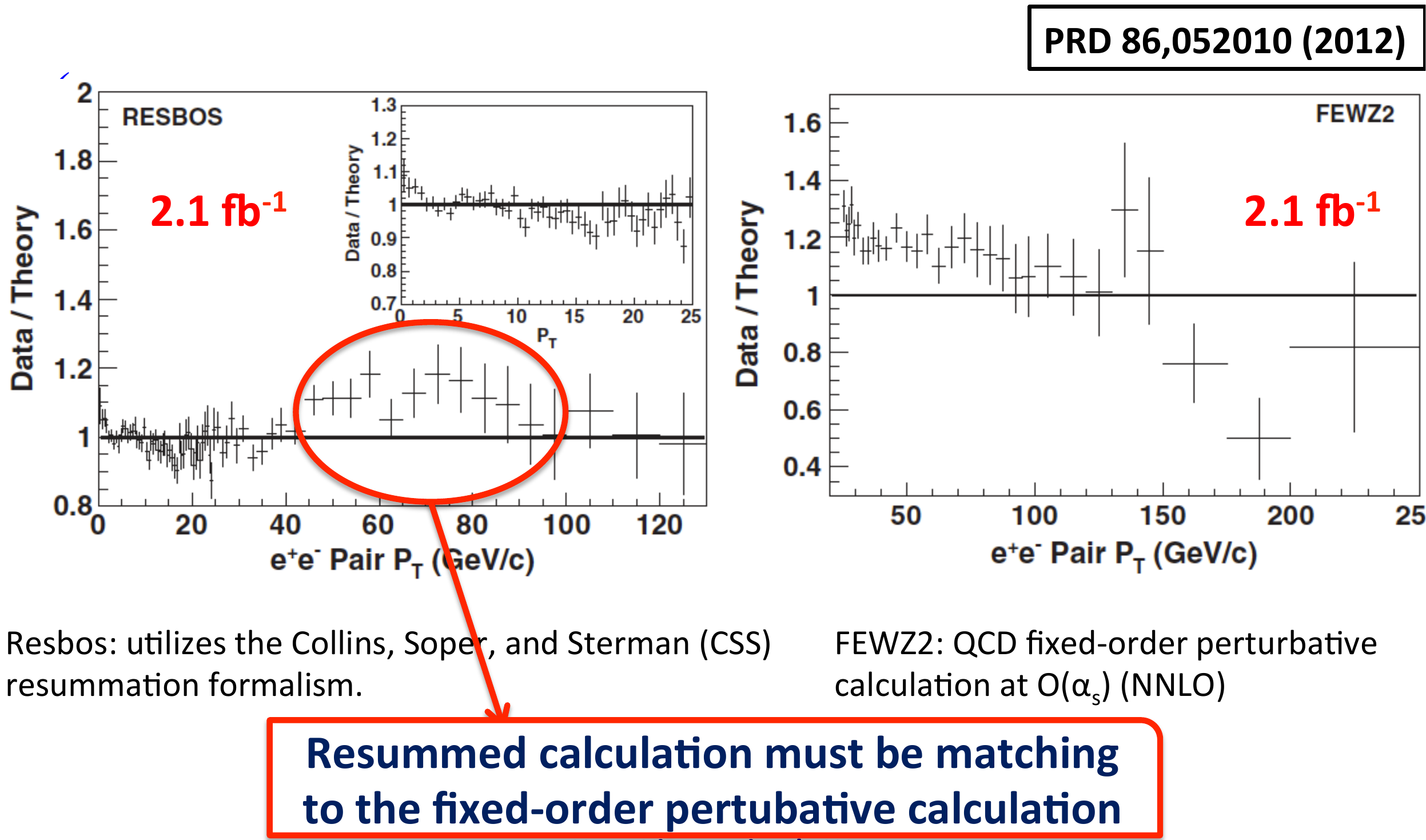}
\end{center}
\caption{\baselineskip=11pt Comparisons of theoretical predictions with the CDF  $Z/\gamma^*$ $p_T$ spectrum in single $Z/\gamma^*$ production with decay to lepton pairs as given in Ref.~\cite{yin}.}
\label{yin-1}
\end{figure}

Now that we have entered the era of precision QCD, 
by which we mean
predictions for QCD processes at the total precision tag of $1\%$ or better,
it is paramount to have rigorous baselines with respect to which to compare 
theoretical results both against one another and against the new LHC precision data as well as for expectations for the future FCC~\cite{fcc} device. 
For example, we have argued in Refs.~\cite{radcor2013,1305-0023,herwiri,qced} that  exact, amplitude-based
resummation allows one to have better than 1\% theoretical precision 
as a realistic goal in such comparisons as those needed in determining the detailed properties of the newly discovered BEH~\cite{EBH} boson~\cite{atlas-cms-2012}, so that one can indeed 
distinguish new physics(NP) from higher order SM processes and can distinguish 
different models of new physics from one another as well. One of the
ingredients in exact amplitude-based resummation is the respective set of hard gluon residuals which determine the order of exactness in 
the respective QCD predictions.
These residuals obtain from exact fixed-order QCD perturbation theory and thus 
any attempt to determine their precision tag necessarily entails determining the respective precision tag of the corresponding fixed-order results. 
Unfortunately, in the current state of the art, even though we have for a process such as single $Z/\gamma^*$ production at LHC(FCC) even the NNLO exact result~\cite{nnlo-gZ},
when one tries to compare the predicted $p_T \;(\text{or}\; \phi^*_\eta)$\footnote{ Here, $\phi^*_\eta$ is a new $p_T$-related variable~\cite{phietastr} used in
some of the comparisons with data and it is defined as follows: $\phi_\eta^*=\tan(\frac{1}{2}(\pi-\Delta\phi))\sin\theta^* \cong \left|\sum \frac{{p_i}_T \sin\phi_i}{Q}\right| +{\cal O}(\frac{{{p_i}_T}^2}{Q^2})$, where $\Delta\phi=\phi_1-\phi_2$ is the azimuthal angle 
between the two leptons which have transverse momenta $\vec{p_i}_T,\; i=1,2,$
and $\theta^*$ is the scattering angle of the dilepton system relative to the beam direction when one boosts to the frame along the beam direction such that the leptons are back to back.} spectrum with the LHC and FNAL data, one sees the type of divergence shown in Figs.~\ref{atlas-hass-1} and ~\ref{yin-1} taken from Refs.~\cite{atlas-hassani,yin}\footnote{Note that in Fig.~\ref{yin-1} the comparisons with RESBOS~\cite{resbos1,resbos2,resbos3}, which realizes the ``CSS'' resummation in Ref.~\cite{css}, show that in the regime where fixed-order result takes over from the resummed terms we see the prediction overshoot the data -- see also the discussion below from Ref.~\cite{rick}.}. 
The ``soft'' limit of the prediction has no reasonable relation to that in the data and even at $p_T \cong 20 GeV$ the so-called exact NNLO result is just useless. Obviously, it will make no sense to talk about the precision tag of such results! Indeed, one of the consequences of the discrepancy in Figs.~\ref{atlas-hass-1},~\ref{yin-1}
is that, until one understands how to fix it, one cannot even be sure of the normalization one gets when one integrates over the theoretical prediction itself.
Indeed, in the precision theory developed and implemented~\cite{yfs-jw} for the LEP physics program, it was in fact true that ``fixing'' such discrepancies changed the normalization. 
\par
\begin{figure}[h]
\begin{center}
%x\epsfig{file=pent-1.eps,width=140mm}
\includegraphics[width=80mm]{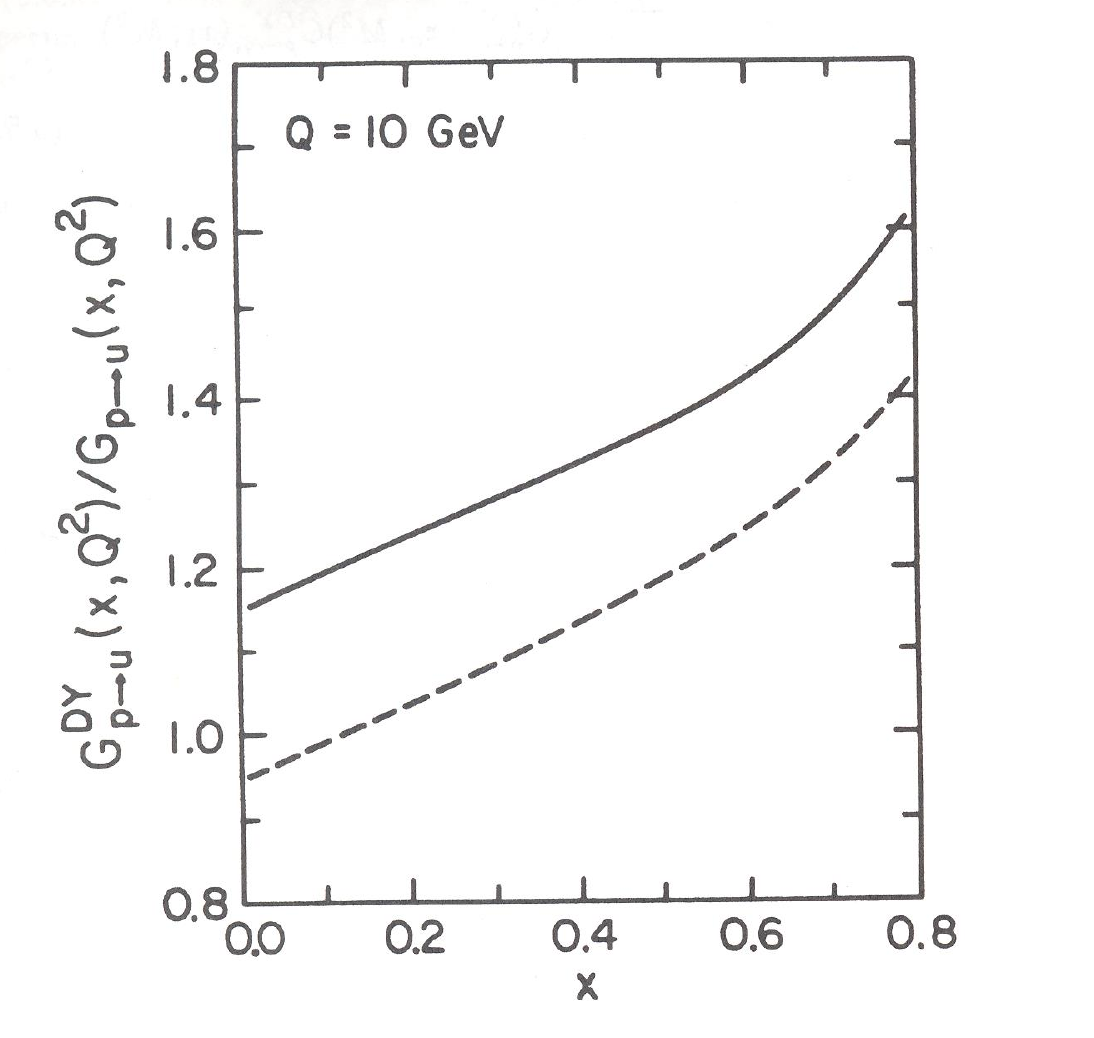}
\end{center}
\caption{\baselineskip=11pt The ratio of the u-quark probability distribution
defined in the Drell-Yan process to that defined from the ${\rm F}_2$ structure function defined in deep inelastic lepton-nucleon scattering as discussed in Ref.~\cite{rick} at $Q= 10$ GeV, where the solid (dashed) curve corresponds to including (excluding) the total cross section in the attendant Drell-Yan distribution.}
\label{fig-rick}
\end{figure}

More precisely, in Refs.~\cite{1305-0023,herwiri}, we have shown 
that the current
realization of the exact amplitude-based resummation 
approach to precision LHC physics as effected by the 
implementation of the IR-improved DGLAP-CS~\cite{dglap,cs} theory~\cite{irdglap1,irdglap2} via HERWIRI1.031~\cite{herwiri} 
in the HERWIG6.5~\cite{herwig} environment improves the agreement between
LHC data on single $Z/\gamma^*$ production in comparison to 
the un-improved predictions. This prepares the stage naturally for 
setting baselines for the respective theoretical precision tags especially
when we focus on the NLO exact, matrix element matched parton shower MC precision issues involved in comparing the predictions of MC@NLO/HERWIRI1.031 and MC@NLO/HERWIG6.5 in the MC@NLO~\cite{mcatnlo} methodology. Here, we define
MC@NLO/A to be the NLO exact, matrix element matched parton shower MC realization of MC A in the MC@NLO methodology. When we try to address these issues,
we are faced with determining the precision of the respective NLO exact matrix element prediction. This latter issue then brings us to the NLO 
version of the type of behavior discussed above for Figs.~\ref{atlas-hass-1},~\ref{yin-1}, which is illustrated in Fig.~\ref{fig-rick} 
and discussed at length in Ref.~\cite{rick} where in its eq.(5.5.30) it is shown that
\begin{equation}
\frac{G^{DY}_{p\rightarrow q}(x,Q^2)}{G_{p\rightarrow q}(x,Q^2)}\operatornamewithlimits{\rightarrow}_{x\rightarrow 1}1+\frac{2\alpha_s(Q^2)}{3\pi}\ln^2(1-x)
\label{rickeq1}
\end{equation}
where $G^{DY}_{p\rightarrow q}$ ($G_{p\rightarrow q}$) is the respective Drell-Yan(DIS) structure function~\cite{rick} in a standard type of notation. 
No observable data, at LHC or the new FCC,
can have this behavior and it calls into question what a precision tag could even mean here?
\par
With an eye toward ``taming'' this what we see in Fig.~\ref{fig-rick} and (\ref{rickeq1}), we revisit our master formula from Refs.~\cite{qced} for our $QED\otimes QCD$ exact resummation theory
\begin{eqnarray}
&d\bar\sigma_{\rm res} = e^{\rm SUM_{IR}(QCED)}
   \sum_{{n,m}=0}^\infty\frac{1}{n!m!}\int\prod_{j_1=1}^n\frac{d^3k_{j_1}}{k_{j_1}} \cr
&\prod_{j_2=1}^m\frac{d^3{k'}_{j_2}}{{k'}_{j_2}}
\int\frac{d^4y}{(2\pi)^4}e^{iy\cdot(p_1+q_1-p_2-q_2-\sum k_{j_1}-\sum {k'}_{j_2})+
D_\rQCED} \cr
&\tilde{\bar\beta}_{n,m}(k_1,\ldots,k_n;k'_1,\ldots,k'_m)\frac{d^3p_2}{p_2^{\,0}}\frac{d^3q_2}{q_2^{\,0}},
%\end{split}
\label{subp15b}
\end{eqnarray}\noindent
where $d\bar\sigma_{\rm res}$ is either the reduced cross section
$d\hat\sigma_{\rm res}$ or the differential rate associated to a
DGLAP-CS~\cite{dglap,cs} kernel involved in the evolution of PDF's and 
where the {\em new} (YFS-style~\cite{yfs-jw,yfs}) {\em non-Abelian} residuals 
$\tilde{\bar\beta}_{n,m}(k_1,\ldots,k_n;k'_1,\ldots,k'_m)$ have $n$ hard gluons and $m$ hard photons and we show the final state with two hard final
partons with momenta $p_2,\; q_2$ specified for a generic $2f$ final state for
definiteness. The infrared functions ${\rm SUM_{IR}(QCED)},\; D_\rQCED\; $ are
defined in Refs.~\cite{qced,irdglap1,irdglap2} as follows:
\begin{eqnarray}
{\rm SUM_{IR}(QCED)}=2\alpha_s\Re B^{nls}_{QCED}+2\alpha_s{\tilde B}^{nls}_{QCED}\cr
D_\rQCED=\int \frac{d^3k}{k^0}\left(e^{-iky}-\theta(K_{max}-k^0)\right){\tilde S}^{nls}_{QCED}
\label{irfns}
\end{eqnarray}
where the dummy parameter $K_{max}$ is such that nothing depends on it.
We have introduced
\begin{eqnarray}
B^{nls}_{QCED} \equiv B^{nls}_{QCD}+\frac{\alpha}{\alpha_s}B^{nls}_{QED},\cr
{\tilde B}^{nls}_{QCED}\equiv {\tilde B}^{nls}_{QCD}+\frac{\alpha}{\alpha_s}{\tilde B}^{nls}_{QED}, \cr
{\tilde S}^{nls}_{QCED}\equiv {\tilde S}^{nls}_{QCD}+{\tilde S}^{nls}_{QED}.
\end{eqnarray} 
The DGLAP-CS synthesization of the infrared functions is denoted
by the superscript $nls$ as explained in Refs.~\cite{dglpsyn,qced,irdglap1,irdglap2} while the infrared functions
$B_A,\; {\tilde B}_A,\; {\tilde S}_A, \; A=QCD,\; QED,$ are given
in Refs.~\cite{yfs-jw,yfs,qced,irdglap1,irdglap2}. 
The exactness of the  
simultaneous resummation of QED and QCD large IR effects that we show here
cannot be emphasized too much. 
\par  
In the interest of pedagogy, we note that,
in the language of Ref.~\cite{gatheral}, 
the exponent ${\rm SUM_{IR}(QCED)}$ sums up to the infinite  order the maximal leading IR singular terms in the cross section for soft emission 
below a dummy parameter $K_{\text{max}}$ and the exponent
$D_\rQCED$ does the same for the regime above $K_{\text{max}}$ so that
(\ref{subp15b}) is independent of $K_{\text{max}}$\footnote{If we want to include more of the
maximal exponentiating terms from the formalism of Ref.~\cite{gatheral} in
the two exponents ${\rm SUM_{IR}(QCED)},\;D_\rQCED$, we may do so with a consequent change in the attendant residuals $\tilde{\bar\beta}_{n,m}$.}.
Exactness order by order in perturbation theory in both 
$\alpha$ and $\alpha_s$ in the presence of these resummed terms, as explained
in Refs.~\cite{qced,irdglap1,irdglap2}, is maintained by iterative 
computation of the residuals $\tilde{\bar\beta}_{n,m}$ to match the attendant 
exact results to all orders in $\alpha$ and $\alpha_s$. In particular, in
our formulation in (\ref{subp15b})
{\it the entire soft gluon phase space is included in the representation -- no part of it 
is dropped}. As it is shown in Refs.~\cite{qced},
the new non-Abelian residuals $\tilde{\bar\beta}_{m,n}$ 
carry a realization 
of rigorous shower/ME matching via their shower subtracted analogs:
in (\ref{subp15b}) we make the replacements
\begin{equation}
\tilde{\bar\beta}_{n,m}\rightarrow \hat{\tilde{\bar\beta}}_{n,m}
\end{equation}
where the $\hat{\tilde{\bar\beta}}_{n,m}$ have had all effects in the showers
associated to the attendant PDF's $\{F_j\}$ removed from them. Here 
we have in mind the standard formula for the  
fully differential representation of a hard LHC(FCC) scattering process:
\begin{equation}
d\sigma =\sum_{i,j}\int dx_1dx_2F_i(x_1)F_j(x_2)d\hat\sigma_{\text{res}}(x_1x_2s),
%       &=\sum_{i,j}\int dx_1dx_2{F'}_i(x_1){F'}_j(x_2)\hat\sigma'(x_1x_2s),
%\end{split}
\label{bscfrla}
\end{equation}
where  
$d\hat\sigma_{\text{res}}$ is given in (\ref{subp15b})
and thus is consistent~\cite{radcor2013,1305-0023,herwiri,qced}
with our achieving a total precision tag of 1\% or better for the total 
theoretical precision of (\ref{bscfrla}). 
\par
For completeness, we also recall the connection between our constructs in the master formula (\ref{subp15b}) and the constructs in the MC@NLO methodology:
We may represent the MC@NLO differential cross section 
via~\cite{mcatnlo} 
\begin{equation}
\begin{split}
d\sigma_{MC@NLO}&=\left[B+V+\int(R_{MC}-C)d\Phi_R\right]d\Phi_B[\Delta_{MC}(0)+\int(R_{MC}/B)\Delta_{MC}(k_T)d\Phi_R]\\
&\qquad\qquad +(R-R_{MC})\Delta_{MC}(k_T)d\Phi_Bd\Phi_R
\label{mcatnlo1}
\end{split}
\end{equation}
where $B$ is Born distribution, $V$ is the regularized virtual contribution,
$C$ is the corresponding counter-term required at exact NLO, $R$ is the respective
exact real emission distribution for exact NLO, $R_{MC}=R_{MC}(P_{AB})$ is the parton shower real emission distribution
so that the Sudakov form factor is 
$$\Delta_{MC}(p_T)=e^{[-\int d\Phi_R \frac{R_{MC}(\Phi_B,\Phi_R)}{B}\theta(k_T(\Phi_B,\Phi_R)-p_T)]},$$
where as usual it describes the respective no-emission probability.
The respective Born and real emission differential phase spaces are denoted by $d\Phi_A, \; A=B,\; R$.
We find it very important {\em still}
to emphasize that the representation of the differential distribution
for MC@NLO in (\ref{mcatnlo1}) illustrates the compensation 
between real and virtual divergent soft effects discussed in the 
Appendices of Refs.~\cite{irdglap1,irdglap2} in establishing the validity of 
(\ref{subp15b}) for QCD. More specifically,
from comparison with (\ref{subp15b}) restricted to its QCD aspect we get the identifications, accurate to ${\cal O}(\alpha_s)$,
\begin{equation}
\begin{split}
\frac{1}{2}\hat{\tilde{\bar\beta}}_{0,0}&= \bar{B}+(\bar{B}/\Delta_{MC}(0))\int(R_{MC}/B)\Delta_{MC}(k_T)d\Phi_R\\
\frac{1}{2}\hat{\tilde{\bar\beta}}_{1,0}&= R-R_{MC}-B\tilde{S}_{QCD}
\label{eq-mcnlo}
\end{split}
\end{equation}
where we defined~\cite{mcatnlo} $$\bar{B}=B(1-2\alpha_s\Re{B_{QCD}})+V+\int(R_{MC}-C)d\Phi_R$$ and we understand 
that the DGLAP-CS kernels in $R_{MC}$ are to be taken as the IR-improved ones
as we derived in Refs.~\cite{irdglap1,irdglap2}. 
Although we have suppressed the superscript $nls$
for simplicity of notation, to avoid double counting of effects the QCD virtual and real infrared functions
$B_{QCD}$ and $\tilde{S}_{QCD}$ are understood to be DGLAP-CS synthesized as explained in Refs.~\cite{qced,irdglap1,irdglap2}. Most importantly, in view of 
(\ref{eq-mcnlo}), we observe that
the way to the extension of frameworks such as MC@NLO to exact higher
orders in $\{\alpha_s,\;\alpha\}$ is open via our $\hat{\tilde{\bar\beta}}_{n,m}$
and will be taken up elsewhere~\cite{elswh}.
\par
We see from the relationship between the hard gluon residuals and the 
exact NLO corrections that a serious study of the theoretical precision
of (\ref{bscfrla}) when it uses the results of (\ref{subp15b}), such as it
is done in Refs.~\cite{herwiri},
necessarily involves the studying of the theoretical 
precision of these exact NLO results and if we have the behavior in 
(\ref{rickeq1}) we do have to ask what would such a study mean in relation
to LHC (or FCC) data? To address this question, we proceed as follows.\par
We recall the well-known representation of the exact NLO 
differential cross section for the Drell-Yan process (we focus on 
the $\gamma^*$ part of the $Z/\gamma^*$ exchange for simplicity of 
presentation, as adding in the effect of the
$Z$ is straightforward and does not affect the analysis here in any 
essential way; similarly, we treat the simple case of one flavor with 
unit charge following Ref.~\cite{guido-mart,humpvn} for the same reason -- 
inserting the proper charges and sums is trivial)
\begin{equation}
\begin{split}
\frac{d\sigma^{DY}}{dQ^2}&=\frac{4\pi\alpha^2}{9sQ^2}\int_0^1\frac{dx_1}{x_1}\int_0^1\frac{dx_2}{x_2}\big\{\left[q^{(1)}(x_1)\bar{q}^{(2)}(x_2)+(1\leftrightarrow 2)\right]\big[\delta(1-z_{12})\\
&\quad + \alpha_s(t)\theta(1-z_{12})(\frac{1}{2\pi}P_{qq}(z_{12})(2t)+f^{DY}_q(z_{12}))\big]\\
&\quad +\left[(q^{(1)}(x_1)+\bar{q}^{(1)}(x_1))G^{(2)}(x_2)+(1\leftrightarrow 2)\right] \\
&\quad \times [\alpha_s(t)\theta(1-z_{12})(\frac{1}{2\pi}P_{qG}(z_{12})t+f^{DY}_G(z_{12}))]\big\}
\end{split}
\label{guido-eq2}
\end{equation}
where $z_{12}=\tau/(x_1x_2), \; \tau=Q^2/s$ in the usual conventions~\cite{guido-mart,rick,humpvn}, the labels $1$ and $2$ refer to the two respective incoming protons
and we follow the generic notation of Refs. ~\cite{rick,guido-mart} here.
The unimproved DGLAP-CS~\cite{dglap,cs} kernels in (\ref{guido-eq2}) are 
well-known as 
\begin{align}
P_{qq}(z)&= C_F \left[\frac{1+z^2}{(1-z)_+} + \frac{3}{2}\delta(1-z)\right],\nonumber\\
P_{qG}(z)&=\frac{1}{2}(z^2+(1-z)^2),
\label{guido-eq3}
\end{align} 
where we define $t=\ln(Q^2/\mu^2)$ following Refs.\cite{guido-mart,rick} so that
$\mu$ is the 't Hooft~\cite{thft-mass} unity of mass.
The scheme dependent hard correction terms are given as follows~\cite{humpvn,guido-mart} if one uses massless quarks and gluons and dimensional regularization, for example:
\begin{equation}
\begin{split}
\alpha_s f^{DY}_G(z)&=\frac{\alpha_s}{2\pi}\frac{1}{2}[(z^2+(1-z)^2)\ln\frac{(1-z)^2}{z}-\frac{3}{2}z^2+z+\frac{3}{2}+2P_{qG}(z)\zeta]\\
\alpha_s f^{DY}_q(z)&=C_F\frac{\alpha_s}{2\pi}\Big[4(1+z^2)\left(\frac{\ln(1-z)}{1-z}\right)_+ -2\frac{1+z^2}{1-z}\ln{z}\\
& \quad+\left(\frac{2\pi^2}{3}-8\right)\delta(1-z)+\frac{2}{C_F}P_{qq}(z)\zeta\Big] 
\end{split}
\label{guido-eq4}
\end{equation}
where we define~\cite{humpvn} $\zeta=-\frac{1}{\epsilon}+C_E-\ln{4\pi}$ for 
$\epsilon=2-n/2$ when $n$ is the dimension of space-time. $C_E$ is Euler-Mascheroni constant.
In the $\overline{\text{MS}}$ scheme, the terms proportional to $\zeta$ are removed 
by mass factorization, which also replaces $\mu$ by $\Lambda$ in $t$ following Ref.~\cite{rick}. This leaves the +-functions in the hard corrections and it is the divergent behavior of these distributions as $z\rightarrow 1$ that produces the attendant unphysical results referenced above. How can we fix this?
\par
We imbed the calculation of the hard correction terms into the master formula
(\ref{subp15b}) restricted to its QCD aspect. This gives the following resummed
version of (\ref{guido-eq2}):
\begin{equation}
\begin{split}
\frac{d\sigma^{DY}_{res}}{dQ^2}&=\frac{4\pi\alpha^2}{9sQ^2}\int_0^1\frac{dx_1}{x_1}\int_0^1\frac{dx_2}{x_2}\big\{\left[q^{(1)}(x_1)\bar{q}^{(2)}(x_2)+(1\leftrightarrow 2)\right]2\gamma_qF_{YFS}(2\gamma_q)(1-z_{12})^{2\gamma_q-1}e^{\delta_q}\\
&\quad \times \theta(1-z_{12})\big[ 1+\gamma_q -7C_F\frac{\alpha_s}{2\pi}+ (1-z_{12})(-1+\frac{1-z_{12}}{2})\\
&\quad +2\gamma_q(-\frac{1-z_{12}}{2}-\frac{z_{12}^2}{4}\ln{z_{12}})\\
&\quad + \alpha_s(t)\frac{(1-z_{12})}{2\gamma_q}f^{DY}_q(z_{12})\big]\\
&\quad +\left[(q^{(1)}(x_1)+\bar{q}^{(1)}(x_1))G^{(2)}(x_2)+(1\leftrightarrow 2)\right] \\
&\quad \times \gamma_G F_{YFS}(\gamma_G)e^{\frac{\delta_G}{2}}[\alpha_s(t)\theta(1-z_{12})\big(\frac{t}{2\pi\gamma_G}(\frac{1}{2}(z_{12}^2(1-z_{12})^{\gamma_G}+(1-z_{12})^2z_{12}^{\gamma_G}))\\
&\quad +f^{DY'}_G(z_{12})/\gamma_G\big)]\big\}
\end{split}
\label{guido-eq5}
\end{equation}
where we have introduced here
\begin{equation}
\begin{split}
\alpha_s f^{DY'}_G(z)&=\frac{\alpha_s}{2\pi}\frac{1}{2}[(z^2(1-z)^{\gamma_G}+(1-z)^2z^{\gamma_G})\ln\frac{(1-z)^2}{z}-\frac{3}{2}z^2(1-z)^{\gamma_G}+z(1-z)^{\gamma_G}\\
&\quad +\frac{3}{4}((1-z)^{\gamma_G}+z^{\gamma_G})],\\
\end{split}
\label{guido-eq6}
\end{equation}
and the following exponents and YFS infrared function, $\FYFS$, already needed for the IR-improvement of DGLAP-CS theory in Refs.~\cite{irdglap1,irdglap2}:
\begin{align}
\gamma_q &= C_F\frac{\alpha_s}{\pi}t=\frac{4C_F}{\beta_0}, \qquad \qquad
\delta_q =\frac{\gamma_q}{2}+\frac{\alpha_sC_F}{\pi}(\frac{\pi^2}{3}-\frac{1}{2}),\nonumber\\
\gamma_G &= C_G\frac{\alpha_s}{\pi}t=\frac{4C_G}{\beta_0}, \qquad \qquad
\delta_G =\frac{\gamma_G}{2}+\frac{\alpha_sC_G}{\pi}(\frac{\pi^2}{3}-\frac{1}{2}),\nonumber\\
\FYFS(\gamma)&=\frac{e^{-{C_E}\gamma}}{\Gamma(1+\gamma)}.
\label{resfn1}
\end{align}
We define $\beta_0=11-\frac{2}{3}n_f$ for $n_f$ active flavors 
in a standard way and $\Gamma(w)$ is Euler's gamma function of the complex variable $w$.
Note that we have mass factorized in (\ref{guido-eq5}) and (\ref{guido-eq6}) 
as indicated above.
It can be seen immediately that the regime at $z_{12}\rightarrow 1$ is now under control in (\ref{guido-eq5}) so that we will no longer have the unphysical
behavior discussed above. This is the main result of this paper.\par
Specifically, instead of the result in (\ref{rickeq1}), we now get the
behavior such that the $\ln^2(1-x)$ on the RHS of (\ref{rickeq1}) is replaced by
$$\frac{2(1-x)^{\gamma_q}\ln(1-x)}{\gamma_q} - \frac{2(1-x)^{\gamma_q}}{\gamma_q^2},$$
and this vanishes for $x\rightarrow 1$. What our result
means that the hard correction now has the possibility to be compared 
{\em exclusively} to the data in a rigorously meaningful way. We take up such matters elsewhere.~\cite{elswh}.
\par
We stress that the parton shower/ME matching formulas in MC@NLO (shown above) and in POWHEG~\cite{powheg} do not remove the IR divergence which we just tamed, as the latter retains the NLO correction with its bad IR limit in the soft regime for $z_{12} \rightarrow 1$ and the former replaces the bad IR behavior of the NLO correction in the soft $z_{12} \rightarrow 1$ limit with that of the parton shower real emission at the same order and it is well known that the respective unimproved parton shower real emission is infrared divergent for $z_{12} \rightarrow 1$ and requires an ad hoc IR cut-off $k_0$-parameter, as we have discussed in Ref.~\cite{herwiri}. {\em No such parameter is needed in our new approach.}\par
To sum up, we have introduced a new approach to hard corrections in perturbative QCD that will allow us to establish the same type of semi-analytical baselines for QCD that we had in Refs.~\cite{yfs-jw} for the higher order 
corrections in the 
Standard Model EW theory.
%\footnote{We note that, while the theory in Refs.~\cite{yfs-jw} was developed for LEP precision theory it has recently been extended~\cite{kkmc422} to EW precision theory for the LHC and FCC.}. 
We look forward to its exploitation in precision LHC and FCC physics scenarios.
In closing, we 
thank Prof. Ignatios Antoniadis for the support and kind 
hospitality of the CERN TH Unit while part of this work was completed.\par

\end{document}